\documentclass[%
 aip,
 amsmath,amssymb,
 reprint,%
]{revtex4-1}

\usepackage{graphicx}
\usepackage{dcolumn}
\usepackage{bm}

\usepackage[utf8]{inputenc}
\usepackage[T1]{fontenc}
\usepackage{mathptmx}
\usepackage{xcolor}

\begin{document}

\preprint{AIP/123-QED}

\title[]{Magnons Parametric Pumping in Bulk Acoustic Waves Resonator}

\author{S.G. Alekseev}
\affiliation{Kotelnikov Institute of Radioengineering and Electronics of Russian Academy of Sciences, Mokhovaya Str.11, build 7, 125009, Moscow, Russia.}

\author{S.E. Dizhur}
\email{sdizhur@phystech.edu.}
\affiliation{Kotelnikov Institute of Radioengineering and Electronics of Russian Academy of Sciences, Mokhovaya Str.11, build 7, 125009, Moscow, Russia.}
\affiliation{Moscow Institute of Physics and Technology (National Research University), Institutsky lane 9, 141700, Dolgoprudny, Moscow region, Russia.}

\author{N.I. Polzikova}%
\email{polz@cplire.ru.}
\affiliation{Kotelnikov Institute of Radioengineering and Electronics of Russian Academy of Sciences, Mokhovaya Str.11, build 7, 125009, Moscow, Russia.}

\author{V.A. Luzanov}
\affiliation{Fryazino branch Kotelnikov Institute of Radioengineering and Electronics of Russian Academy of Sciences, Vvedenskogo Square, 1, 141400, Fryazino, Moscow region, Russia.}

\author{A.O. Raevskiy}
\affiliation{Fryazino branch Kotelnikov Institute of Radioengineering and Electronics of Russian Academy of Sciences, Vvedenskogo Square, 1, 141400, Fryazino, Moscow region, Russia.}

\author{A.P. Orlov}
\affiliation{Kotelnikov Institute of Radioengineering and Electronics of Russian Academy of Sciences, Mokhovaya Str.11, build 7, 125009, Moscow, Russia.}

\author{V.A. Kotov}
\affiliation{Kotelnikov Institute of Radioengineering and Electronics of Russian Academy of Sciences, Mokhovaya Str.11, build 7, 125009, Moscow, Russia.}

\author{S.A. Nikitov}
\affiliation{Kotelnikov Institute of Radioengineering and Electronics of Russian Academy of Sciences, Mokhovaya Str.11, build 7, 125009, Moscow, Russia.}
\affiliation{Moscow Institute of Physics and Technology (National Research University), Institutsky lane 9, 141700, Dolgoprudny, Moscow region, Russia.}

\date{\today}

\begin{abstract}
We report on the experimental observation of excitation and detection of parametric spin waves and spin currents in the bulk acoustic wave resonator. 
The hybrid resonator consists of ZnO piezoelectric film, yttrium iron garnet (YIG) films on gallium gadolinium garnet substrate, and a heavy metal Pt layer. 
Shear bulk acoustic waves are electrically excited in the ZnO layer due to piezoeffect at the resonant frequencies of the resonator. 
The magnetoelastic interaction in the YIG film emerges magnons  (spin waves) excitation by acoustic waves either on resonator’s eigenfrequencies or the half-value frequencies at supercritical power. We investigate acoustic pumping of magnons at the half-value frequencies and  acoustic spin pumping from parametric  magnons, using the inverse spin Hall effect in the Pt layer.
The constant electric voltage in the Pt layer, depending on the frequency, the magnetic field, and the pump power, was systematically studied. 
We explain the low threshold obtained ($\sim 0.4$ mW) by the high efficiency of electric power transmission into the acoustic wave in the resonator. 
\end{abstract}

\maketitle

The studies in the field of magnonics or magnon spintronics direct to employ microwave spin waves (SW) as main carriers of information data transmission and processing.\cite{Khitun2010,Chumak2017,Nikitov_2015,DEMIDOV2017}
Therefore, the problem of interaction between SW (or their quanta -- magnons) and other condensed matter excitations (phonons, electrons, etc.) is of great importance.
The numerous manifestations of  SW - acoustic wave (AW) interaction in micro- and nanoscale structures, especially the conversion of magnetic energy into elastic and vice versa, are of interest both for practical applications and fundamental physics.\cite{Delsing2019,Bozhko2020,Hashimoto2018,Latcham2019}

The elastic  excitation of SW can be carried out without  the alternating magnetic fields, which substantially reduce ohmic losses in comparison with  the inductive current-driven excitation.\cite{Cherepov2014} So the magnon-phonon interaction  is very promising for the development of low energy SW logic circuits and memory elements.

Linear excitation of acoustically driven spin waves (ADSW) was demonstrated in various hybrid magnon-phonon structures containing piezoelectric (PE) and ferromagnetic layers.\cite{Bas2019,Gowtham2015,Labanowski2017,Azovtsev2019}
To excite ADSW, the close contact between the layers is unnecessary, and a high-Q acoustic medium can separate the layers with neither PE nor ferromagnetic properties.\cite{An2020,Polzikova1998,Polzikova2013}
Strain-induced magnon effects, such as SW generation, propagation, and amplification in magnetoelectric structures with close contacts of PE and thin ferromagnetic films also gained a lot of attention recently.\cite{Khitun2010, Sadovnikov2018, Cherepov2014}

Ferromagnetic (at microwave frequencies) yttrium iron garnet (Y$_3$Fe$_5$O$_{12}$ - YIG) grown onto gadolinium gallium garnet (GGG) substrate has extremely low losses of spin waves and acoustic waves, and their high conversion efficiency. 
It makes YIG-GGG the perfect candidate for magnon spintronics devices. Thus, investigations of acoustically driven spin waves (ADSW) in heterostructures with YIG-GGG are now quite a hot topic in scientific research.


A widely used spin pumping method is now implemented to study SW dynamics in micro- or nanoheterostructures.\cite{Manuilov2015,Hahn2013,Fukami2016,Noack2019}
Spin pumping (SP) is a transform of the spin angular momentum of magnons into a spin current at the ferromagnetic-nonmagnetic metal interface.\cite{Tserkovnyak2002} 
The electrical detection of SW results from the spin to charge current conversion in heavy metal like Pt due to the inverse spin Hall effect (ISHE).\cite{Saitoh2006} 
We will refer to this method as SP/ISHE further. 

Spin pumping from linearly excited ADSWs with a surface AW was experimentally exhibited.\cite{Weiler2012,Uchida2011,Bhuktare2019}
In Refs.~\onlinecite{Polzikova2016,Polzikova2018AIP,Polzikova2018,POLZIKOVA2019}, we demonstrated the piezoelectric ADSWs excitation and their SP/ISHE detection in the spintronic hypersonic High overtone Bulk Acoustic wave Resonator (HBAR) with  ZnO/YIG/GGG/YIG/Pt structure. 
In a similar HBAR structure, the influence of the parametric acoustic pumping on the propagation of SW excited via traditional microstrip antennas was studied.\cite{Chowdhury2017}

The excitation of short wavelength parametric magnons is currently studied in connection with the tendency to reduce the size of magnonic structures. Therefore, the source of low energy parametric magnons is highly relevant.
Apparently, purely acoustic parametric excitation of SW in HBAR has not been reported previously.
The SP/ISHE method for detecting parametric ADSWs up to the present day was considered only theoretically.\cite{KESHTGAR2014}

In this work, we report the observation of the parametric ADSW excitation in a spintronic HBAR with YIG/Pt structure used previously in  Ref.~\onlinecite{Polzikova2018AIP} for the linear ADSW excitation.
We studied the frequency and magnetic field dependences of the ISHE constant voltage signal under the resonator's excitation in the GHz frequency range. 
The signal from the parametric ADSW was observed when: 
i) the pump frequency coincided with one of the HBAR resonant frequencies, 
ii) half of the pump frequency coincided with one of the SW frequencies for a given magnetic field, 
iii) the pump power exceeded a threshold, depending on the field. 

\begin{figure}[h]
\includegraphics[width=\columnwidth]{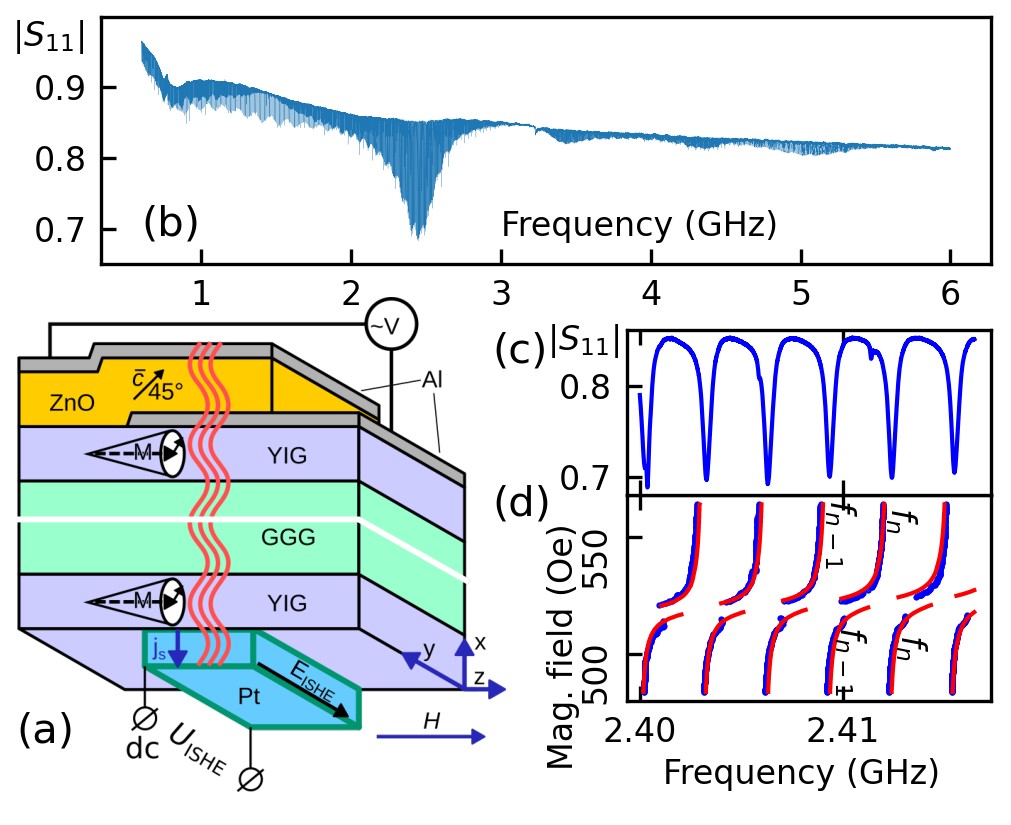}
\caption{\label{fig:1} (a) Schematic of hybrid HBAR with the Pt stripe for acoustic SP. (b),(c) Reflectivity $|S_{11}(f)|$ in the frequency domain at zero magnetic field. 
(d) Positions of HBAR resonant frequency $f_{n}$ in the $(f, H)$-plane: 
the blue curves -- experiment, the red curves -- calculation according to the theory in Ref.~\onlinecite{POLZIKOVA2019}.}
\end{figure}


The scheme of the experimental HBAR is shown in Fig.~\ref{fig:1}(a). The resonator was fabricated on the basis of a ready-made YIG-GGG-YIG wafer consisting of a single-crystal $500$ $\mu$m thick GGG (111) plate and two $30$ $\mu$m thick YIG films grown on both sides of the plate by liquid-phase epitaxy. The YIG films were doped with La and Ga and characterized by a saturation magnetization of $4\pi M_{\mathrm{s}} \sim 800-900$~G. The use of  these films provides a number of advantages  with respect to pure YIG films ($4\pi M_{\mathrm{s}} \sim 1750$~G), in particular, the SW excitation  at lower frequencies, comparable to the frequencies of AW generated by PE transducers.\cite{Polzikova2016,Polzikova2018}
Note that the presence of the upper YIG film is not necessary for the effects considered but is dictated by the YIG’s growth technology.


A PE transducer consisting of ZnO film sandwiched between thin-film Al electrodes was deposited on one side of the YIG-GGG-YIG structure. 
The electrodes were patterned by photolithography and had an overlap with the aperture $a = 170$ $\mu$m. 
The textured ZnO film produced by rf magnetron sputtering has the inclined $\mathbf{c}$-axis for the excitation of the primary shear AWs with elastic displacement in the YIG films plane.\cite{Luzanov2018} 

Figure~\ref{fig:1}(b) shows the HBAR spectrum – frequency dependence of the microwave voltage reflection coefficient  $|S_{11}(f)|$ in the absence of an external magnetic field. 
The enlarged fragment in Fig.~\ref{fig:1}(c) is given for the region of the most efficient thickness modes excitation, in which all measurements were then performed.
The presence of  dips in $|S_{11}(f)|$ (sharp reduction of reflectivity) indicates the excitation of AW, since they carry out the energy. The difference between adjacent dips is $f_n - f_{n-1} \approx 3.1$ MHz, and corresponds to the excitation of thickness extensional shear AW.
Then, the structure was placed in an external tangential magnetic field $\mathbf{H} \| \mathbf{z}$, which magnetized the YIG films to saturation $4\pi M_{\mathrm{s}} = 845$ G if $H > H_{\mathrm{s}} \sim 10$ – $20$ Oe, where $H_{\mathrm{s}}$ is the saturation field.

In YIG films, ADSWs are excited due to the interaction between the magnetic and elastic subsystems.\cite{Kittel1958} 
For their detection by the SP/ISHE method, a $12$ nm thin Pt layer was deposited on the bottom YIG film's surface underneath the PE transducer aperture. 
Spin pumping from ADSW results in the time-averaged spin current $\mathbf{j}_s \propto \mathbf{n} \langle (\mathbf{M} \times \partial \mathbf{M}/\partial t ) \rangle_z = j_s\mathbf{n}$ polarized along $\mathbf{z}$ and flowing from YIG into Pt parallel to the normal $\mathbf{n} \| \mathbf{x}$. 
The ISHE in Pt leads to an electrostatic field $\mathbf{E}_{\mathrm{ISHE}} \propto - (\mathbf{j}_s \times \mathbf{z}$) in the $\mathbf{y}$-direction, and the dc voltage $U_{\mathrm{ISHE}} = - a(\mathbf{E}_{\mathrm{ISHE}}\cdot \mathbf{y}$) between Pt stripe’s ends.\cite{Saitoh2006}
Frequency dependencies of $U_{\mathrm{ISHE}}(f)$ and $|S_{11}(f)|$ were simultaneously measured at a fixed magnetic field, varied in the range $H_{\mathrm{s}} \leq | H | \leq 1$ kOe. 

Figure~\ref{fig:1}(d) displays the resonance frequencies' rearrangement due to linear ADSW excitation under magnetoelastic resonance (MER) at the field $H_{\mathrm{MER}} \approx 520$ - $525$ Oe. 
Similar to our previous works (see Refs.~\onlinecite{Polzikova2016,Polzikova2018AIP,Polzikova2018,POLZIKOVA2019}), a voltage $U_{\mathrm{ISHE}} (f, H)$ from the excitation of linear ADSW was also observed in the vicinity of $H_{\mathrm{MER}}$ that will be discussed later.

\begin{figure}[h]
\includegraphics[width=\columnwidth]{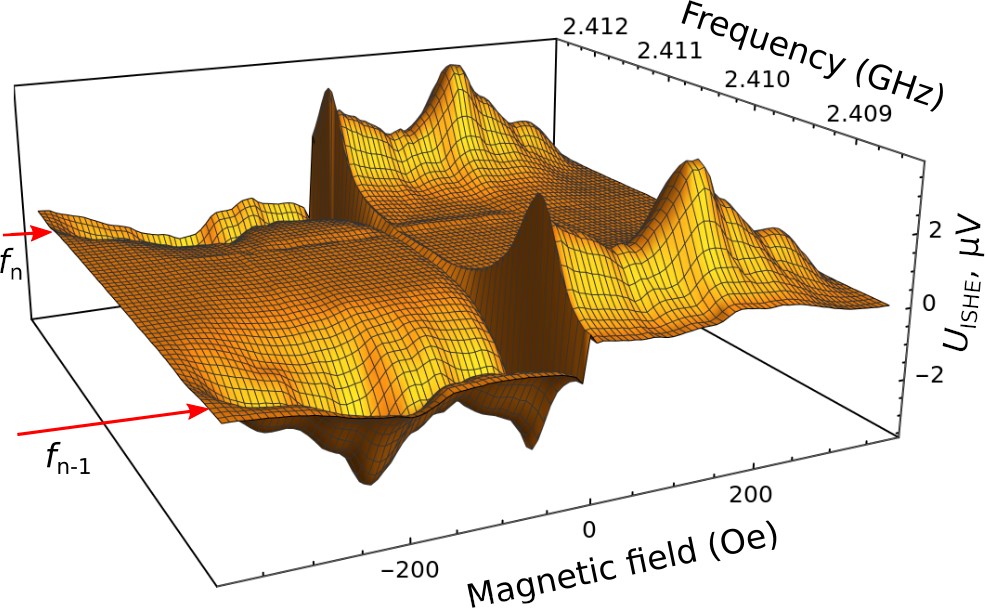}
\caption{\label{fig:2_3D} 3D dependence of $U_{\mathrm{ISHE}}(f, H)$ at the applied power $P = 9$ mW.}
\end{figure}

While the magnetic field increases above the range shown in Fig.~\ref{fig:1}(d), the frequency spectrum does not experience a noticeable effect, and there is no dc voltage signal. 
When the field decreases below the range, the spectrum also remains almost unchanged, but the voltage signal $U_{\mathrm{ISHE}} (f, H)$ is clearly detected, as shown in Fig.~\ref{fig:2_3D}. 
The maxima of $U_{\mathrm{ISHE}} (f, H)$ correlate with the positions of the HBAR resonance frequencies $f_n$ and $f_{n-1}$, indicated by arrows.  
Note, when the field direction is reversed, the voltage changes the sign, which is typical for the ISHE symmetry. 
Narrow region of magnetic fields $|H| \sim 2$-$3$ Oe $< H_{\mathrm{s}}$, where the voltage changes the sign during magnetization reversal, is of undoubted interest, but outside of the scope of this research and will be discussed in the further work.

Let us consider in details the magnetic field dependence of the voltage at a fixed frequency
in an infinite medium
$f_{\mathrm{p}} = f_n$ in comparison with the dependencies of the characteristic SW frequencies 
$f_{\mathrm{SW}}(q,H) = \gamma\sqrt{H_{\mathrm{eff}}(q)(H_{\mathrm{eff}}(q) + 4\pi M_{\mathrm{s}} \sin^2\theta)} = \sqrt{f_H(q)\times (f_H(q)+f_M\sin^2\theta)}$.
Here, $\gamma = 2.8$ MHz/Oe, $H_{\mathrm{eff}}(q,H)= H + 4\pi M_{\mathrm{s}}(\lambda q)^2$, $\lambda \sim  10^{-6}$ cm is the exchange length, and $\theta$ is the angle between the wave vector $\mathbf{q}$ and $\mathbf{H}$. 

\begin{figure}[h]
\includegraphics[width=\columnwidth]{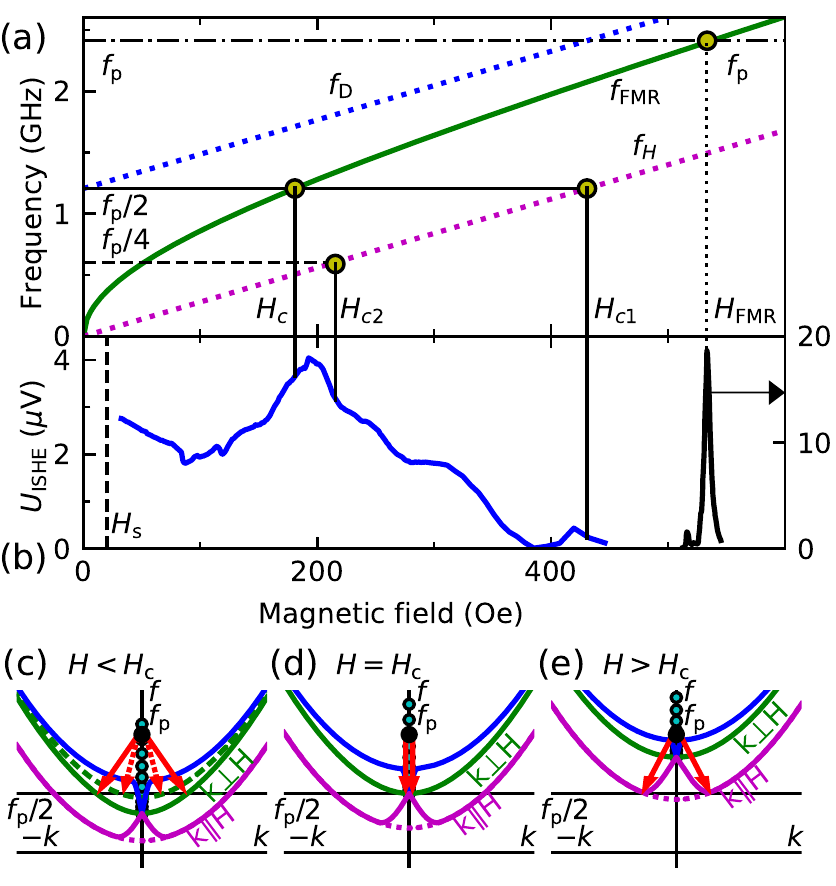}
\caption{\label{fig:3} (a) Dependences of SW frequencies at $q = 0$ in an infinite medium} on the magnetic field: $f = f_H(0$) (dotted magenta) at $\theta = 0$; $f = f_{\mathrm{FMR}}$ (green) at $\theta = 90^{\circ}$.
The dotted blue line represents the frequency
$f = f_{\mathrm{D}} = f_H(0) + f_M / 2$ -- upper boundary of dipole (non-exchange) surface magnetostatic waves. 
(b) Experimental field dependencies $U_{\mathrm{ISHE}}(H)$ at the pump frequency $f_{\mathrm{p}} = f_n = 2.412$ GHz. 
(c)-(e) Dispersion diagrams $f_{\mathrm{SW}}(k)$ and pumping schemes at different fields. 
The blue circles denote the SW resonance frequencies.
\end{figure}

Figure~\ref{fig:3}(a) shows the calculated field dependences of the frequency limits of the SW spectra at $q = 0$ with parallel ($\theta  = 0$, $f = f_H(0,H)$) and perpendicular ($\theta  = 90^{\circ}$,   $f = f_{\mathrm{FMR}}(H)$) directions of $\mathbf{q}$.
Here $f_{\mathrm{FMR}} = f_{\mathrm{SW}}(0,H)$ is the ferromagnetic resonance frequency. 
The horizontal lines mark the resonant frequency HBAR $f_{\mathrm{p}} = f_n = 2.412$ GHz, and it’s sub-harmonics $f_{\mathrm{p}} /2$ and $f_{\mathrm{p}} /4$. 
The dependency $U_{\mathrm{ISHE}}(H)$ (blue line) measured with the step of $1$ Oe and corresponding to the cross-section of the surface $U_{\mathrm{ISHE}} (f, H)$ at $f = f_n$ (see Fig.~\ref{fig:2_3D}) is shown in Fig.~\ref{fig:3}(b).
It can be seen from Fig.~\ref{fig:3}(a) and Fig.~\ref{fig:3}(b) that the field of the maximum voltage $H = 193$ Oe is close to the critical field $H_c = 184$ Oe, at which $f_{\mathrm{FMR}}(H_c) = f_{\mathrm{p}}/2$. 
It indicates that $U_{\mathrm{ISHE}}(H_c)$ originates from the detection of parametric SW with $q = 0$. 
The voltage signal is observed in the field range below of $432$ Oe. 
This field correlates with the value $H_{c1} = f_{\mathrm{p}} /(2\gamma)$ in Fig.~\ref{fig:3}a.
Above this field, the excitation of any parametric SW is impossible; therefore, the voltage signal is suppressed. 
However, at higher fields, the linear excitation of ADSW results in the peak of $U_{\mathrm{ISHE}}(f_n, H_{\mathrm{FMR}})$, presented by the black curve in  Fig.~\ref{fig:3}(b). 
Here $H_{\mathrm{FMR}} \approx 534$ Oe is found from the relationship $f_n  = f_{\mathrm{FMR}}(H)$.

Consider now a range of wavenumbers where parametric ADSWs can be excited. 
Figures~\ref{fig:3}(c)-\ref{fig:3}(e) schematically show the dispersion $f_{\mathrm{SW}}(\pm k)$ of SW propagating in the film plane at an angle $\theta$ to the magnetic field direction: $\theta  = 0$ (magenta curves), and $\theta  = 90^{\circ}$ (green curves).
Here $k = |\mathbf{k}| = |\mathbf{q}_{\|}|$ is the in-plane component of the wave vector.
Blue curves represent the dispersions of surface SW.
Figure~\ref{fig:3}(d) displays the dispersion diagram at the field $H = H_c$. 
The vertical arrows show the process mentioned above of the creation of two parametric magnons with $k = q = 0$ and frequency $f_{\mathrm{p}}/2$.

In the fields $H_c < H < H_{c1}$, the spectrum of parametric SW is a set of the backward volume SW, propagating in-plane at angles $90^{\circ}  > \theta  \geq  0$. 
In Fig~\ref{fig:3}(e), the arrows demonstrate the pumping scheme of parametric exchange SWs at $H \approx H_{c1}$. 
The SWs are excited with sufficiently large wavenumbers $k_0$ and zero group velocity.
For a film of thickness $d$, the minimal SW frequency increases slightly: $(f_{\mathrm{min}}(k_0) - f_H(0)) / f_H(0) \sim  \lambda/d \sim 10^{-3}$ (see Ref.~\onlinecite{Polzikova1984}). 
On the contrary, the wavenumbers of minima turn out to be substantially different from zero: $k_0 \sim 1 / (\lambda d)^{1/2} \sim 10^4$ cm$^{-1}$.

The field region $H < H_c$ corresponds to a dipole-exchange SW propagating perpendicular to the field ($\theta = 90^{\circ}$). 
Figure~\ref{fig:3}(c) illustrates the parametric generation of exchange SWs of two types: propagating mainly in the film plane with large $k$ and $q_{\perp} \ll k$ (solid arrows), and standing modes of the SW thickness resonance with $k \approx 0$, $q_{\perp} \gg k$ (dashed arrows). 
Here $q_{\perp} = \pi s/d$ is the wavenumber components along the film thickness and $s = 1,2,\ldots$ are the SW resonance numbers. 
We suppose that the change of the type of parametrically excited SW evokes the observed non-monotonic dependence of $U_{\mathrm{ISHE}}(H)$ at $H \sim 100$ Oe (see Fig.~\ref{fig:3}(b)).
The total wavenumber $q=|\mathbf{k} + \mathbf{q}_{\perp}|$ can be estimated from the relation $(q\lambda)^2 = (H_c - H)/4\pi M_{\mathrm{s}}$, which provides for the maximum wavenumber $q \sim 10^5$ cm$^{-1}$ at $H = H_{\mathrm{s}}$.
As one can see from Figure~\ref{fig:3}(a) the frequencies $f_n/2$
in this case are close to the upper frequency limit $f_{\mathrm{D}} = f_H(0) + f_M / 2$
for dipole (non-exchange) surface SW in the plate.

\begin{figure}[h]
\includegraphics[width=\columnwidth]{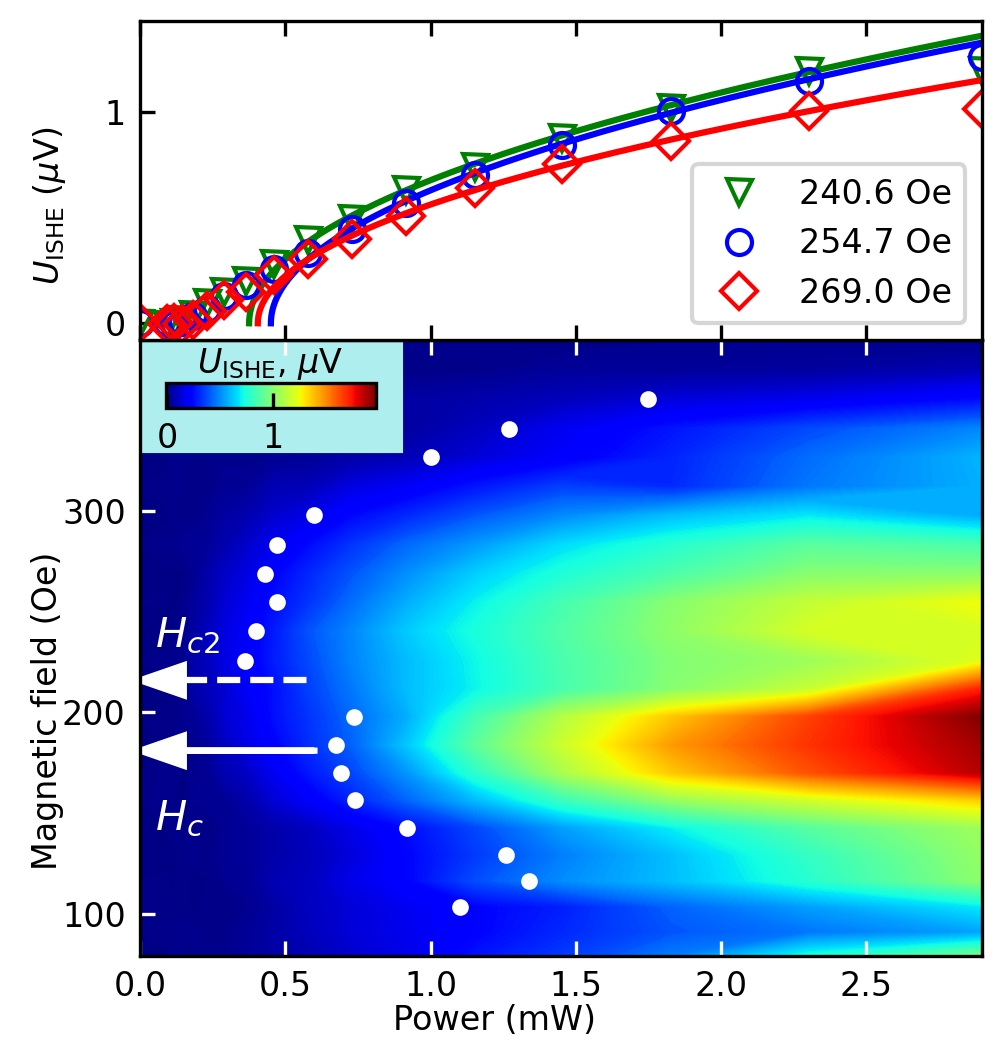}
\caption{\label{fig:4UISHE} (a) The dependences $U_{\mathrm{ISHE}}(P)$ at different magnetic fields.
The symbols -- experiment results, the lines -- the fitting with the relation $U_{\mathrm{ISHE}} \propto (P-P_{\mathrm{th}}(H))^{1/2}$. 
(b) Color-density plot represents $U_{\mathrm{ISHE}}(H,P)$ measured at $f_{\mathrm{p}} = f_n = 2.412$ GHz.
White circles show $P_{\mathrm{th}}(H)$.}
\end{figure}

The experiments described above
were conducted at a fixed power level of $9$ mW. 
Next, we study the voltage dependence $U_{\mathrm{ISHE}}(P)$ at lower power levels up to $3$ mW.
Figure~\ref{fig:4UISHE}(a) shows the $U_{\mathrm{ISHE}}(P)$ for three values of the magnetic field $H > H_c$. 
The data are fitted by $U_{\mathrm{ISHE}}(P,H) \propto (P-P_{\mathrm{th}}(H))^{1/2}$, where the fitting parameter $P_{\mathrm{th}}(H) \approx 0.4$ mW can be interpreted as a threshold power. 
Similar power dependences are specific to the concentration of the magnon dominant group, excited near the threshold of parametric instability $P_{\mathrm{th}}(H)$.\cite{Zakharov1975} 
The results of the $U_{\mathrm{ISHE}}(P,H)$ measurements over a wide magnetic field range are shown in the color-coded plot in Fig.~\ref{fig:4UISHE}(b). 
The white circles show $P_{\mathrm{th}}(H)$ found from the fitting.

The solid arrow marks the critical field $H_c$. 
According to the previous consideration (Fig.~\ref{fig:3}(b) and \ref{fig:3}(d)), a parametric process is possible with the creation of SW at a frequency $f_{\mathrm{p}}/2$ and with $q = k = 0$. 
As it can be seen, the local minimum of the threshold power $P_{\mathrm{th}}(H)$ is also located in this field. 
The dashed arrow marks the field $H_{c2} = H_{c1}/2 = f_{\mathrm{p}}/4\gamma$. 
It corresponds to the upper limit on $H$ for the possible decays of parametric magnons with the frequency $f_{\mathrm{p}}/2$ into two secondary parametric ones at a frequency $f_{\mathrm{p}}/4$. 
A jump in $P_{\mathrm{th}}(H)$ behavior is precisely at $H = H_{c2}$, and it can be attributed to an increase in the magnetic relaxation parameter $\Delta H_q$ due to the contribution of secondary parametric processes. 

The mechanisms of magnetoelastic parametric excitation of magnons from pumping by AW with different polarizations were considered in Refs.~\onlinecite{Matthews1964, HAAS1966, FILIMONOV2004, Kraimia2020, Lisenkov2019}. 
Following Ref.~\onlinecite{HAAS1966}, for pumping by the shear AW propagating perpendicular to the field, we obtained the expression for the threshold acoustic power $P_S = 0.5\pi a^2 C_S V_S (M_{\mathrm{s}} \Delta H_q/B_2)^2 \approx 20\ldots 160$ $\mu$W.
Here $C_S = 7.64\times 10^{11}$ dyn/cm$^2$ and $V_S = 3.85\times 10^5$ cm/s are the elastic modulus and velocity for shear AW, respectively; $B_2 = (5\ldots 7) \times 10^6$ erg/cm$^3$ is the magnetoelastic constant, $\Delta H_q = 0.25$ \ldots $0.5$ Oe. 
The relationship of this power with the power applied to the electrodes can be found from  $P_S = P_{\mathrm{th}}\Delta|S_{11}|^2$, where $\Delta |S_{11}|^2$ is the value of the dip $|S_{11}(f)|^2$ in the vicinity of the resonant frequency $f_n$.\cite{Weiler2012, Polzikova2016} 
From Fig.~\ref{fig:1}(c), we found $\Delta|S_{11}|^2 \approx 0.2$; hence $P_{\mathrm{th}} = 0.1 \ldots 0.8$\,mW, which is consistent with the experiment (see Fig.~\ref{fig:4UISHE}).
Thus, based on the preceding, we conclude that the voltage observed results in the detection of parametric ADSW with the SP/ISHE method.

{\ }\\

In summary, the acoustic generation of parametric spin waves -- ADSW -- was observed in the resonator of bulk AW containing YIG/GGG/YIG/Pt structure.
Parametric ADSW were detected using the spin pumping from YIG to Pt and ISHE in Pt.
The dependence of the voltage on the pump power shows that the minimum thresholds are quite small, which is due to the excitation efficiency of the high-Q hybrid HBAR by the PE transducer strictly at the resonant frequencies.
It should be noted the Q-factor of our resonator at 2.4 GHz was $Q\approx 4\times 10^3$, that is far from the possible values ($Q>10^5$) which can be obtained in HBAR at this frequency. Thus increasing Q-factor and excitation frequency as well as decreasing the ferromagnetic film thickness makes it possible to excite magnons with wavenumbers  exceeding $q\sim 10^5$cm$^{-1}$ reported in this work.
{\ }\\

This work was carried out in the framework of the State task 0030-2019-0013 "Spintronics" and with partial support of the RFBR (Project No. 20-07-01075). The work of SAN is supported by the Government of the Russian Federation for the state support of scientific research conducted under the guidance of leading scientists in Russian higher-education institutions, research institutions, and state research centers of the Russian Federation (Project No. 2019-220-07-9114).

{\ }\\
The authors declare no conflicts of interest.

\section*{Data availability}

The data that support the findings of this study are available from the corresponding authors upon reasonable request.

\section*{References}

\bibliography{Magnons}

\end{document}